\documentclass[pra,superscriptaddress,twocolumn,showpacs,floatfix]{revtex4}
\usepackage{graphicx}
\usepackage{epstopdf}
\usepackage{amsmath}
\usepackage{color}
\usepackage{amsfonts}
\usepackage{subfigure}
\usepackage{bm}

\def\bra#1{\mathinner{\langle{#1}|}}
\def\ket#1{\mathinner{|{#1}\rangle}}
\def\braket#1{\mathinner{\langle{#1}\rangle}}

\begin{document}

\title{Wave Function Renormalization Effects in Resonantly Enhanced Tunneling}

\author{N. L\"orch}
\affiliation{Institut f\"ur Theoretische Physik, Universit\"at
Heidelberg, Philosophenweg 19, 69120 Heidelberg, Germany}

 \author{F. V. Pepe}
\affiliation{Dipartimento di Fisica and MECENAS, Universit\`a di
Bari, I-70126 Bari, Italy} \affiliation{INFN, Sezione di Bari,
I-70126 Bari, Italy}

\author{H. Lignier}
 \affiliation{Laboratoire Aim\'e Cotton, Universit\'e Paris-Sud, Batiment 505, Campus d'Orsay, F-91405 Orsay Cedex, France}

\author{D. Ciampini}
\affiliation{CNISM-Pisa,  Dipartimento di Fisica, Universit\`{a}
di Pisa, Lgo Pontecorvo 3, 56127 Pisa, Italy}
 \affiliation{INO-CNR, Dipartimento di Fisica, Universit\`{a} di
Pisa, Lgo Pontecorvo 3, I-56127 Pisa,Italy}

\author{R. Mannella}
\affiliation{CNISM-Pisa, Dipartimento di Fisica,  Universit\`{a}
di Pisa, Lgo Pontecorvo 3, 56127 Pisa, Italy}

\author{O. Morsch}
 \affiliation{INO-CNR, Dipartimento di Fisica, Universit\`{a} di
Pisa, Lgo Pontecorvo 3, I-56127 Pisa,Italy}

\author{E. Arimondo}
\affiliation{CNISM-Pisa, Dipartimento di Fisica, Universit\`{a} di
Pisa, Lgo Pontecorvo 3, 56127 Pisa, Italy}
 \affiliation{INO-CNR, Dipartimento di Fisica, Universit\`{a} di
Pisa, Lgo Pontecorvo 3, I-56127 Pisa,Italy}

\author{P. Facchi}
\affiliation{Dipartimento di Matematica and MECENAS, Universit\`a
di Bari, I-70125 Bari, Italy}
\affiliation{INFN, Sezione di Bari, I-70126 Bari, Italy}

 \author{G. Florio}
\affiliation{Dipartimento di Fisica and MECENAS, Universit\`a di
Bari, I-70126 Bari, Italy}
\affiliation{INFN, Sezione di Bari, I-70126 Bari, Italy}

 \author{S. Pascazio}
\affiliation{Dipartimento di Fisica and MECENAS, Universit\`a di
Bari, I-70126 Bari, Italy}
\affiliation{INFN, Sezione di Bari, I-70126 Bari, Italy}

\author{S. Wimberger}
\affiliation{Institut f\"ur Theoretische Physik, Universit\"at Heidelberg, Philosophenweg 19, 69120 Heidelberg, Germany}
\affiliation{Center for Quantum Dynamics, Universit\"at Heidelberg, 69120 Heidelberg, Germany}

\begin{abstract}

We study the time evolution of ultra-cold atoms in an accelerated optical
lattice. For a Bose-Einstein condensate with a
narrow quasi-momentum distribution in a shallow optical lattice the decay of the survival
probability in the ground band has a step-like structure. In this
regime we establish a connection between the wave function renormalization parameter $Z$
introduced in [Phys.\ Rev.\ Lett.\ {\bf 86}, 2699
(2001)] to characterize non-exponential decay and the phenomenon
of resonantly enhanced tunneling, where the decay rate is peaked
for particular values of the lattice depth and the accelerating
force.
\end{abstract}

\pacs{03.65.Xp, 03.75.Lm}

\date{\today}

\maketitle

\section{Introduction}
Resonantly enhanced tunneling (RET) is a quantum effect in which
the probability for the tunneling of a particle between two potential
wells is increased when the quantized energies of the initial and
final states of the process coincide. In spite of the fundamental
nature of this effect~\cite{bohm} and its practical
interest~\cite{chang}, it has been difficult to observe it
experimentally in solid state structures. Since the 1970s, much
progress has been made in constructing solid state systems such as
superlattices~\cite{chang_app,esaki_review,Glutsch2004} and
quantum wells~\cite{wagner93} which enable the controlled
observation of RET~\cite{leo03}.

In recent years, ultra-cold atoms in optical
lattices~\cite{grynberg,rev1}, arising from the interference
pattern of two or more intersecting laser beams, have been
increasingly used to simulate solid state systems \cite{Bloch2005,rev1,rev2}.
Optical lattices are easy to realize in the laboratory, and the parameters
of the resulting one-, two- or three-dimensional periodic
potentials (the lattice spacing and the potential depth) can be
perfectly controlled both statically and dynamically.
In the papers \cite{Sias:2007,Zenesini:2008}, a Bose-Einstein condensate
(BEC) in accelerated optical lattice potentials was used to study the
phenomenon of RET. In a tilted periodic potential, atoms can
escape by tunneling to the continuum via higher-lying levels.
Within the RET process the tunneling of atoms out of a tilted
lattice is resonantly enhanced when the energy difference between
lattice wells matches the distance between the energy levels in
the wells.

The atomic temporal evolution is described by the survival
probability, starting from an initial state prepared in the
ground band of the lattice. At long interaction times, after
several tunneling processes, the survival probability is characterized by an exponential decay rate with a constant
tunneling probability for each Bloch period \cite{Glueck2002}. Such a decay
was examined in different theoretical
analyses~\cite{wagner93,Glutsch2004,Glueck2002} and measured in
experimental investigations with ultra-cold atoms
\cite{Sias:2007,Zenesini:2008,Zenesini:2009,Tayebirad:2010}.
In this study we scrutinize the time behavior of the tunneling probability and use its remarkable features at short and intermediate times in order to extract information about wave-function renormalization effects.

The key quantity in this context is the probability that the system investigated ``survives" in a given state (or a set of states, such as a band of a lattice).
In this article we shall deal with survival probabilities whose behavior is complex and difficult to analyze. See for example the experimental results of ref.~\cite{Zenesini:2009} and the
Figs.\ \ref{simulation} and \ref{SurvivalProbability} in the following, which display the survival probability of a cloud of ultra-cold atoms in the ground band of an accelerated optical lattice. Clearly, one can properly speak of the ``decay" associated with an unstable system (the atoms tend to leak out of the accelerated lattice), but the time evolution can display oscillations or even plateaus. (As we shall see, the latter are easily understood in terms of the initial atomic state.)

General theoretical considerations show that the (adiabatic) survival probability of an unstable system can often be written as
\begin{equation}
P(t) = Z \exp\left(-\gamma t\right) + \textrm{additional contributions},
\label{Pexp}
\end{equation}
where $\gamma$ is the decay rate, which can be computed by the Fermi golden rule, and the parameter
$Z$, representing the extrapolation of the asymptotic decay law back to $t=0$, is related to wave-function
renormalization.
Law (\ref{Pexp}) is valid both in quantum mechanics \cite{textbook00,textbook0} and quantum field theory \cite{textbook1,textbook2}, and $Z$ can be smaller or larger than unity \cite{Pascazio}.
Typically, the additional contributions in (\ref{Pexp}) dominate both at short and long times, where the exponential decay law is superseded by a quadratic \cite{Misra,Pascazio2,temprevi} and a power law \cite{Khalfin57}, respectively.
They are therefore crucial in order to cancel the exponential in these time domains. However, they can play a key role in a much more general context, such as the RET phenomenon to be investigated in this article.

The pioneering experiments performed in Texas, with Landau-Zener transitions in cold atoms,
checked the existence of the short-time quadratic behavior
\cite{Wilkinson:1997} and the transition \cite{Fischer:2001} from the quantum Zeno effect \cite{Misra} to the
anti- or inverse-Zeno effect \cite{Lane,FP,KK}, through a sequence of properly
tailored quantum measurements.

With the arrival of Bose-Einstein condensates the experimental
resolution has advanced even further as compared to cold atoms.
While cold atoms can have a momentum distribution on the order of
a Brillouin zone or more, a very narrow distribution (much smaller
than a Brillouin zone) is achievable with BECs.  Even the steplike
structure of the survival probability occuring for shallow lattice
depth can be resolved with great precision \cite{Zenesini:2009,Tayebirad:2010}. It is
in this regime of shallow lattices and short jump times
\cite{Vitanov} where the yet unobserved link of RET and the
initial deviation from exponential decay is most striking.
This work is devoted to the study of these effects. The choice of a different initial atomic state, with a well defined momentum, will enable us to observe a more complicated temporal structure.
We shall therefore scrutinize the time evolution in order to unveil an exponential regime and introduce the $Z$ parameter in our RET framework.

The paper is organized as follows.
We briefly sum up previous results on RET and the quantum Zeno
effect in Section \ref{ret-zeno}. We then analyze the dynamics in the
tilted lattice in Section \ref{study}, and show in Section
\ref{intuitive}, the main part of this article, how the two
phenomena arise as interference effects.
Section \ref{exptconf} reports
experimental results for the wave-function renormalization parameter $Z$ in the case of a Bose-Einstein condensate in an
accelerated optical lattice, and also a comparison with the experimental
configuration by Wilkinson {\em et al.}~\cite{Wilkinson:1997}.
Section \ref{concl} concludes our work.

\section{Landau-Zener and resonantly enhanced tunneling}
\label{ret-zeno}
A Landau-Zener (LZ) transition takes place in a system with a
time-dependent Hamiltonian, in which the spectrum, as a function of
a control parameter (here time $t$), is characterized by the presence
of an avoided crossing \cite{Landau,Zener,Stueckelberg,Majorana,Jones}. A LZ transition is described
by the following two-level Hamiltonian
\begin{equation}\label{landauzener}
H_{\mathrm{LZ}}(t)= \left( \begin{array}{cc} \alpha t & \delta E/2
\\ \delta E/2 & -\alpha t \end{array} \right),
\end{equation}
written in a suitable basis, known as {\it diabatic basis}. The expectation values of Eq.\
(\ref{landauzener}) on the two states of the  basis depend
linearly on time and cross at $t=0$. On the other hand, the
coupling $\delta E/2$ between the  states is constant. The
diagonalization of Eq.\ (\ref{landauzener}) yields the eigenvalues
\begin{equation}
\label{eig-adiabatic}
E_{\pm}=\pm\sqrt{(\alpha t)^2 + \left(\frac{\delta E}{2} \right)^2}.
\end{equation}
The eigenbasis of $H_{\mathrm{LZ}}(t)$ is called the
{\it adiabatic basis}. At $t\rightarrow-\infty$ the adiabatic energy
levels of Eq.\ (\ref{eig-adiabatic}) are infinitely separated, and no
transition between them occurs. The distance between the levels
decreases towards the avoided crossing at $t=0$, and then increases
again until, at $t\rightarrow\infty$, the separation becomes again
infinite. If the system is prepared at $t\rightarrow-\infty$ in
one of the adiabatic eigenstates, the probability that the system undergoes a transition at
$t\rightarrow\infty$ towards the other adiabatic eigenstate reads \cite{Zener}
\begin{equation}
\label{plz}
P_{\mathrm{LZ}}=\exp\left(-\frac{\pi(\delta E)^2}{4\hbar\alpha}   \right).
\end{equation}

\begin{figure}
\begin{center}
\includegraphics[width=0.5\textwidth]{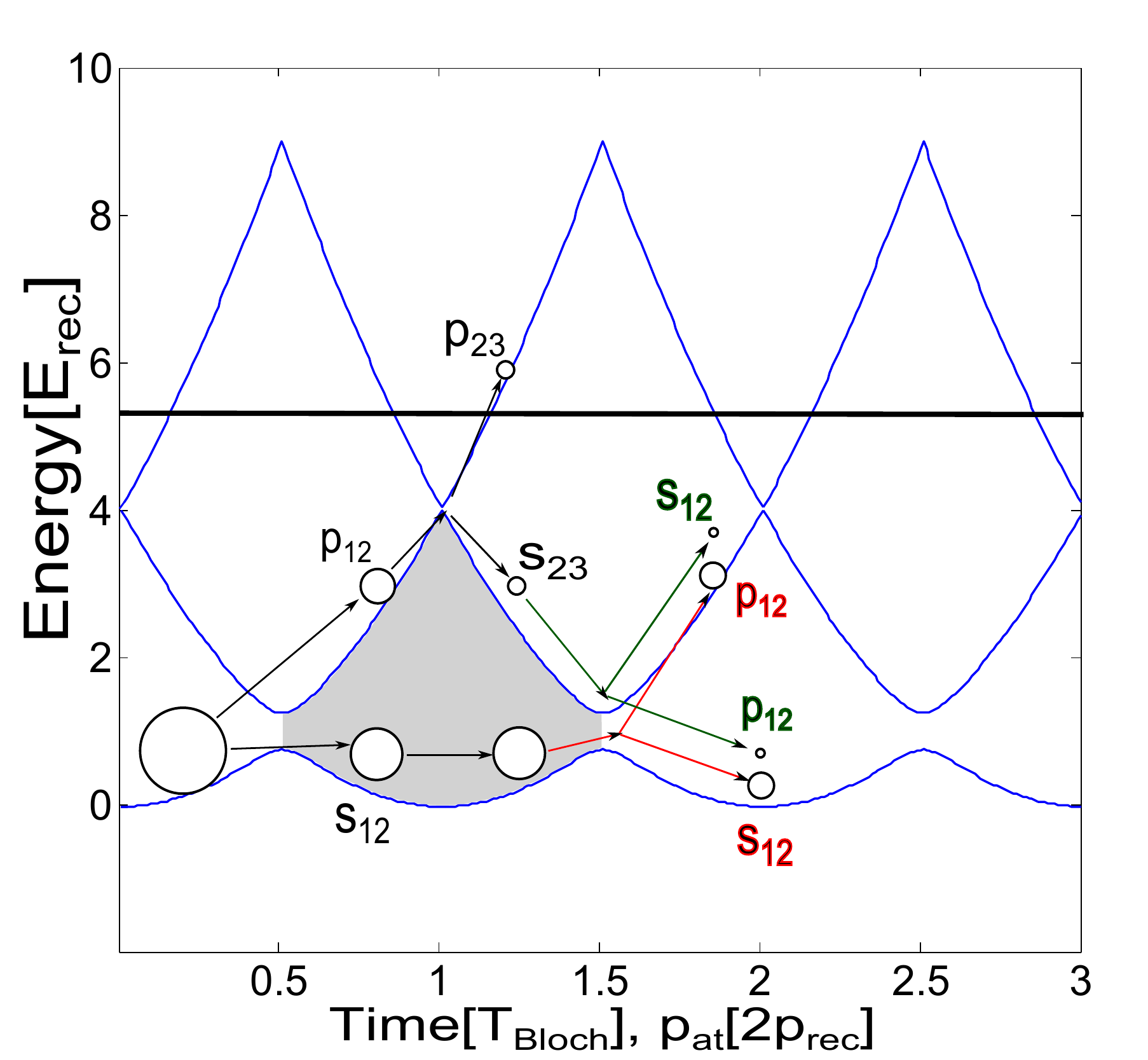}
\end{center}
\caption{ (color online) Energy diagram for a particle in a periodic potential vs  either the time, in units of Bloch time $T_{\rm B}$ defined in~(\ref{Tbloch}),
and the  atomic momentum $p_{\mathrm{at}}/2p_{\mathrm{rec}}$ where $p_{\mathrm{rec}}= \hbar \pi/d_L$ units.
Under the application of an external force, the quasi-momentum increases with time and at the avoided crossing between two bands, at the edge of the Brillouin zone, where $p_{\mathrm{at}}=p_{\mathrm{rec}}$, the condensate tunnels with a
probability amplitude $p_{ij}$ and survives with an amplitude $s_{ij}$, defined in
Eq.\ \eqref{evored} below. Between two avoided crossings of the lowest two bands a relative
phase $\phi$, defined in Eq.~\eqref{phi12},  is acquired, which is graphically displayed in the
figure as a grey area. The final survival probability at
a given time is the sum over all possible routes, just like a path
integral in momentum space.}
\label{anschaulich}
\end{figure}

A particle in a shallow periodic potential, subjected to an
external force, is an example of a physical system in which a LZ
process can be observed. In this case, the diabatic basis is
represented by the momentum eigenstates. As schematized in Fig.~\ref{anschaulich},
if the system is
initially prepared in the lowest band, with a very peaked momentum
distribution around $p=0$, it will evolve towards the edge of the
first Brillouin zone, where the distance between the first and the
second band is minimal and transitions are more likely to occur,
and then evolves back to the bottom of the first band. The
transition probability towards the second band in this process can
be approximated by $P_{\mathrm{LZ}}$, but discrepancies can arise
due to the differences between the idealized case, leading to the
LZ formula (\ref{plz}), and the real physical situation.
Indeed, the periodicity of the lattice implies that the
aforementioned process occurs in a {\it finite} time, and that in
the initial and final states the adiabatic levels are not
infinitely separated. The corrections to the LZ transition
probability due to the finite duration of the process are
discussed in \cite{Holthaus,Tayebirad:2010}.

Other corrections to Eq.\ (\ref{plz}) should be considered if the
lattice is not shallow. In this case, couplings to higher
momentum states play an important role and a two-level description
is not a good approximation anymore.

Moreover, there is another kind of deviation from LZ, which will
be the main object of our analysis. Since Eq.\ (\ref{plz}) is obtained
under the hypothesis that only one of the two adiabatic
eigenstates is initially populated, it is not valid anymore if both
states are populated. These deviations can be relevant even if
one of the initial populations is very close to zero, since their
order is {\it square root} of the smaller population, as will
be discussed in the following.
In a periodic potential, tilted by an external force $F$, the
probability that a wave packet initially prepared in the first
band jumps to the second band corresponds to the LZ prediction~\eqref{plz}
only if the second band is empty. A small population in the second
band gives rise to oscillations around $P_{\mathrm{LZ}}$.

Finally, the transition probability is enhanced by a large
factor with respect to the LZ prediction if the energy difference
$Fd_L\Delta i$ between two potential wells ($d_L$ being the lattice
spacing and $d_L\Delta i$ the distance between the wells) matches
the average band gap of the non-tilted system (RET).
One expects that in a RET process from the first to the second band, the asymptotic
regime will only be reached after a transient period. Indeed, while the first transition
occurs when the second band is strictly empty (and thus the tunneling event closely follows the LZ prediction),
further RET transitions will occur periodically in time and, starting from the second tunneling process,
interference effects due to the finite
population amplitude in the second band will start to play an important role, modifying the time evolution in an important way.

The analysis of the following two sections will endeavor to take all these effects into account. We shall build up an effective model, whose validity will be tested for rather diverse ranges of the parameters, and compared to experimental results finally in section \ref{exptconf}.

\section{Dynamics of interband tunneling}
\label{study}

In our analysis we are interested in describing the RET process
from the first to the second band of a Bose-Einstein condensate
loaded into an optical lattice. It will be assumed that almost
all the particles of the system are in the condensate, so that the
system is described by a single-particle wave function $\psi(x,t)$
\cite{Stringari}. Moreover, we will consider the condensate
dilute enough so that the interaction between particles can be neglected.
This implies that the wave function of the system obeys a linear
Schr\"odinger equation. Nonlinear effects have been studied in the
RET regime in previous works \cite{Sias:2007,Zenesini:2008,AW2011,Wim2005,WimZ1,WimZ2,WimZ3,WimZ4}.

The experimental condition is that of an
accelerating one-dimensional optical lattice, with constant
acceleration $a$. In the rest frame of the lattice, a particle of
mass $m$ sitting in the lattice is subjected to an external force $F=ma$,
and thus the time-independent Hamiltonian of the system in this
frame of reference reads
\begin{equation}\label{hamilt}
H=-\frac{\hbar^2}{2m}\frac{\partial^2}{\partial x^2}
+\frac{V}{2}\cos \left( \frac{2\pi x}{d_L} \right)-Fx\equiv
H_0-Fx,
\end{equation}
where $V$ is the lattice depth and the lattice period $d_L$
is  half wavelength  of the counterpropagating lasers. $H_0$ represents
the ``unperturbed'' Hamiltonian, whose eigenstates are the Bloch
functions
\begin{align}\label{b}
& \phi_{\alpha,k}(x)=\mathrm{e}^{ikx}\chi_{\alpha,k}(x), \\
& \chi_{\alpha,k}(x+d_L)=\chi_{\alpha,k}(x), \\
& H_0 \phi_{\alpha,k}(x)= E_{\alpha}(k)\phi_{\alpha,k}(x),
\end{align}
with $\alpha=1,2,3,\dots$ the band index and $k$ the
quasi-momentum, ranging in the first Brillouin zone
$\mathcal{B}=\{k|-\pi/d_L\leq k\leq \pi/d_L\}$. The dynamics of the
system depends on two dimensionless parameters \cite{Tayebirad:2010},
related to lattice depth and external force:
\begin{equation}
\label{rescaling}
V_0=\frac{V}{E_{\mathrm{rec}}},  F_0=\frac{F
d_L}{E_{\mathrm{rec}}}, \text{ with }
E_{\mathrm{rec}}=\frac{\hbar^2}{2m}\left( \frac{\pi}{d_L}
\right)^2,
\end{equation}
where $m$ is the mass of the atoms. Applying the unitary transformation
\begin{equation}
\tilde{\psi}(x,t)=\exp(-iFxt/\hbar)\psi(x,t)
\end{equation}
restores the
translational invariance of the Hamiltonian, at the expense of an
explicit time dependence:
\begin{equation}
\label{Hamiltonianequation}
\tilde{H}(t)=\frac{1}{2m}\left(-i\hbar\frac{\partial}{\partial
x}+Ft \right)^2 +\frac{V}{2}\cos \left( \frac{2\pi x}{d_L}
\right).
\end{equation}
Rewriting this Hamiltonian in the momentum basis as in \cite{Tayebirad:2010} establishes the relation to the Landau-Zener Hamiltonian introduced in Eq.~\eqref{landauzener}: 
To calculate the time evolution of any momentum eigenstate, 
we only need the  Hamiltonian $H_{k_0}$ acting on the subspace with a given quasi-momentum $k_0$, as there is no transition between states with different $k_0$
\begin{equation}
  H_{k_0} = \frac{1}{2m} \left(
    \begin{array}{ccccc}
      \ddots & & & & 0 \\
        & \hbar^2(k-\frac{2\pi}{d_L})^2 &m V/2 & &  \\       
        & mV/2 & (\hbar k)^2 & mV/2 &  \\  
        & & mV/2 & \hbar^2(k+\frac{2\pi}{d_L})^2 &  \\       
        0 & & & & \ddots \\
    \end{array}
  \right)\ ,
  \label{eqno8}
\end{equation}
where $k=k(t)=k_0+Ft/\hbar$.
This  Hamiltonian (\ref {eqno8}) leads to a very accurate numerical
solution of the Schr\"odinger equation. For small $V$ on the scale of $E_{\mathrm{rec}}$ its dynamics is well described by successive Landau-Zener transitions, 
occuring whenever two diagonal terms in $H_{k_0}$ become degenerate. We will use this approximation to obtain analytical results. In Fig.~\ref{anschaulich} 
the relevant transitions are depicted graphically.


We first examine an adiabatic approximation of the dynamics generated by the Hamiltonian (\ref{hamilt}),
yielding no transition between bands ({\it single-band
approximation}), which will highlight the time periodicity of the
system and the phase differences between bands. We shall then introduce
an effective coupling between the low-lying bands that enables one to
obtain transition rates. The
adiabatic approximation is consistent if $Fd_L\lesssim V$, namely if  $F_0 \lesssim V_0$ in Eq.\ (\ref{rescaling}).

The initial state will be assumed to be highly peaked around a single
quasi-momentum value $k_0$, that is, the width of the initial
quasi-momentum distribution will be taken to be much smaller than the width
$2\pi/d_L$ of the first Brillouin zone $\mathcal{B}$.
In this situation, it can be proved \cite{Jones,Holthaus} that in
the adiabatic single-band approximation the average quasi-momentum
evolves semiclassically, so that at time $t$
\begin{equation}\label{kclass}
k(t)=k_0+\frac{F t}{\hbar},
\end{equation}
with negligible spread in the quasi-momentum distribution occurring
during the evolution. This yields Bloch oscillations
in a tilted lattice with a Bloch period
\begin{equation}\label{Tbloch}
T_{\rm B}=\frac{2\pi\hbar}{F
d_L}=\frac{\hbar}{E_{\mathrm{rec}}}\frac{2\pi}{F_0}.
\end{equation} The initial state analyzed here has a well
defined initial momentum (in $\mathcal{B}$), but can be
distributed among different bands. At the end of each Bloch
period, the amplitude in band $\alpha$ acquires the following
phase with respect to the amplitude in band $\beta$
\begin{equation}\label{Blochphase}
\Delta\varphi_{\alpha\beta}=  -\frac{1}{\hbar} \langle
E_{\alpha}(k)-E_{\beta}(k) \rangle T_{\rm B} = -\frac{2\pi}{F_0} \langle
\epsilon_{\alpha}(k)-\epsilon_{\beta}(k) \rangle ,
\end{equation}
where $\langle \dots\rangle$ denotes the average over
$\mathcal{B}$
and $\epsilon_{\gamma}(k)$ is the energy of the
state with quasi-momentum $k$ in band $\gamma$ in units $E_{\mathrm{rec}}$.

\begin{figure}
\begin{center}
\includegraphics[width=0.45\textwidth]{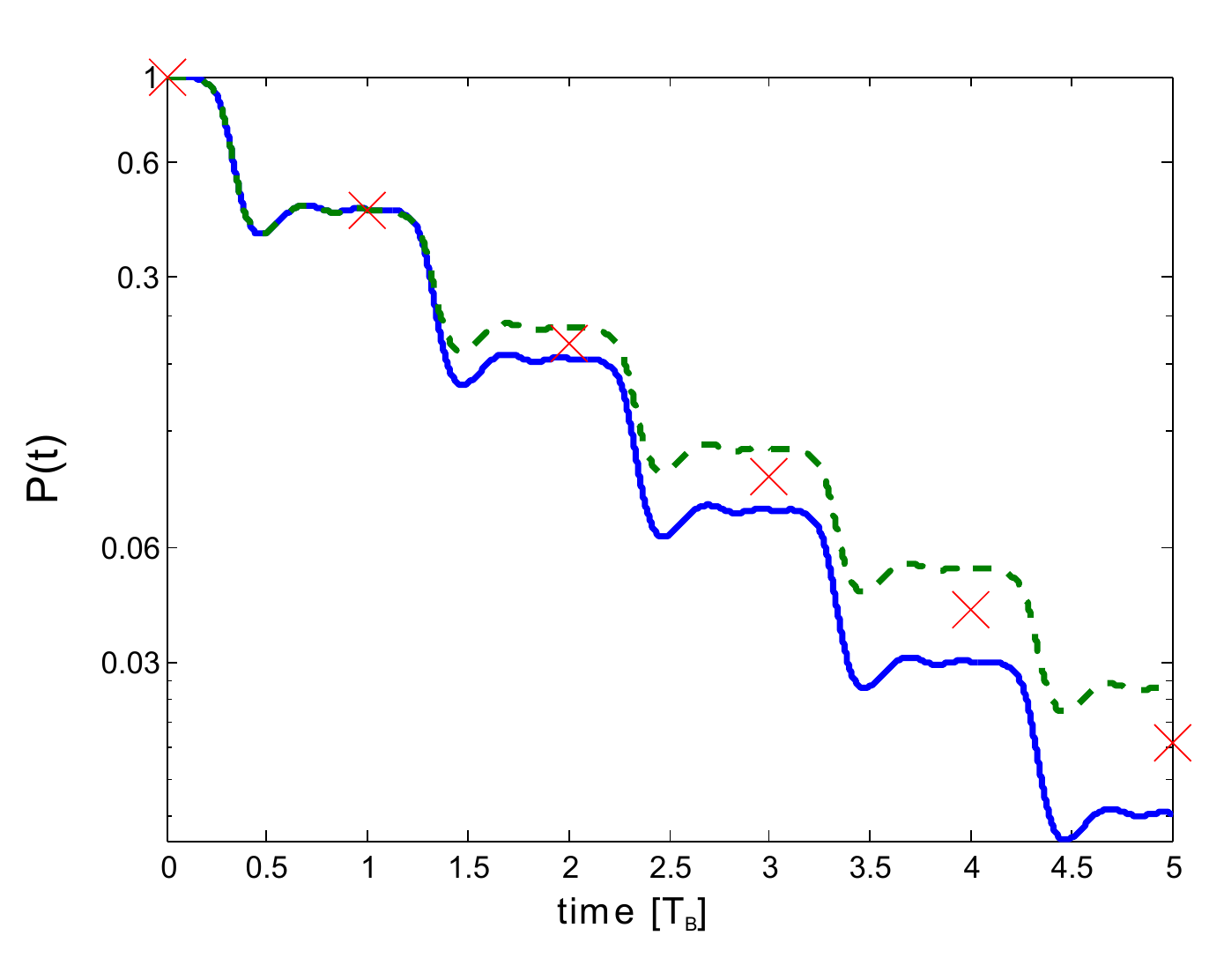}
\end{center}
\begin{center}
\vspace{-6mm}\hspace{3.8mm}\includegraphics[width=0.45\textwidth]{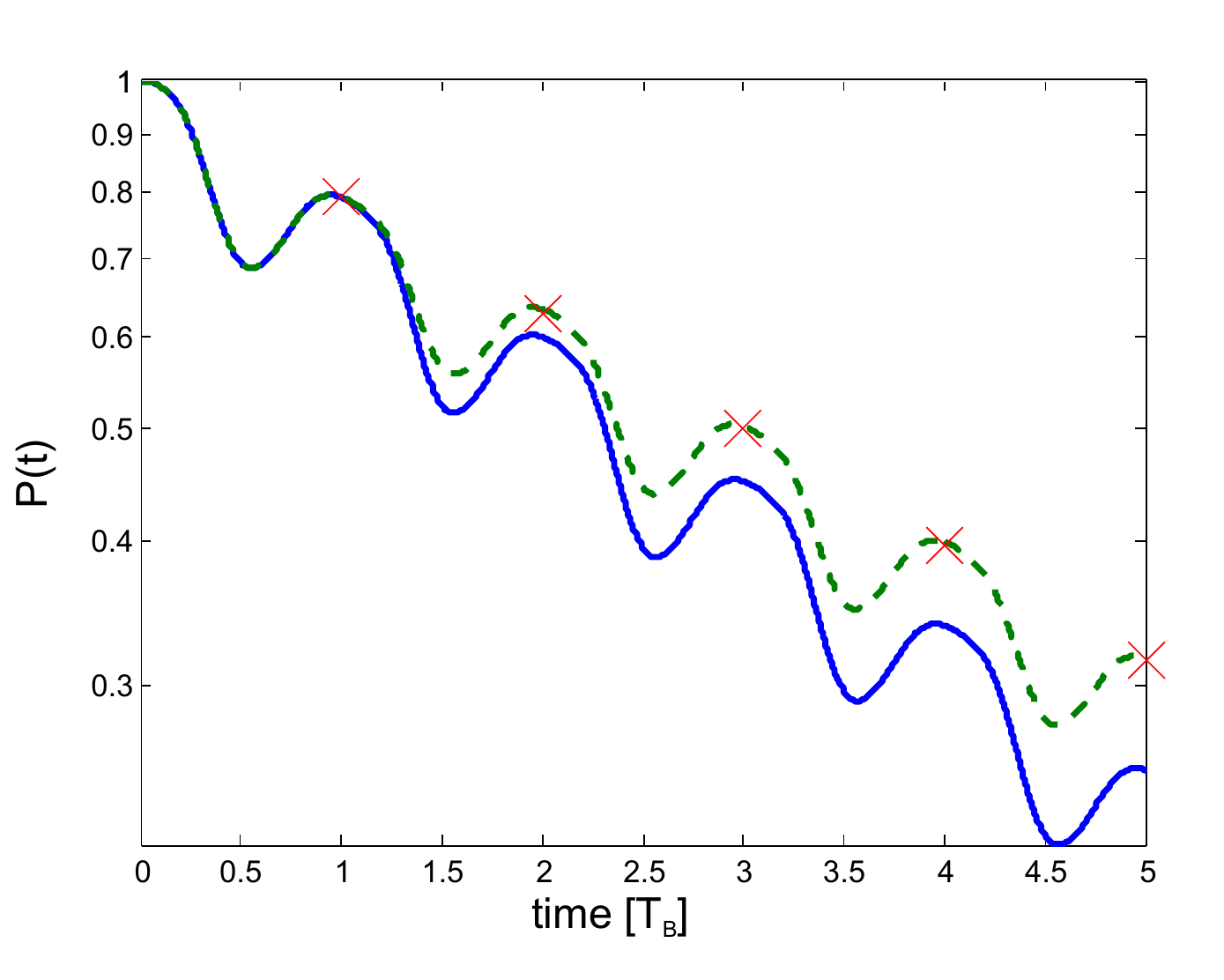}
\end{center}
\caption{(color online) Adiabatic survival probability  in the lowest band
$P(t)$ vs. time, obtained by numerically solving the Schr\"odinger equation for the atomic evolution under the Hamiltonian of Eq.\ (\ref{Hamiltonianequation}). The initial state has initial quasi-momentum $0.2 p_{\textrm{rec}}$ in the Brillouin zone and
a negligible quasi-momentum width. Upper panel: $V_0=1.5, \phi \simeq 4 \pi$;
lower panel: $V_0=4.5, \phi \simeq 2 \pi$.
The solid (blue) line represents the unperturbed time evolution, the dashed (green) line
the time evolution with the  phase change discussed in Sec.\ \ref{phasecontrol} after each Bloch
period. The red crosses are an extrapolation of the first time step to the following periods. Plateaus manifest for $V_0=1.5$ (upper panel). The validity of this picture is discussed in the text and in  Appendix\ \ref{shallowlatt}.}
\label{simulation}
\end{figure}

We now analyze inter-band
transitions through an effective model. We focus on the experimental parameters of the Pisa setup \cite{Zenesini:2009,Tayebirad:2010} and model transitions from the first to the second band.
In the parameter regime of shallow lattices there is numerical and
experimental evidence of a step-like structure of the adiabatic survival
probability $P(t)$ \cite{Zenesini:2009} in the first band. If the initial state is peaked
around $k=0$ and lies in the first band, the survival probability
is characterized by steep drops around times $t=T_{\rm B}(n+1/2)$ with
$n$ integer, and flat plateaus between these times
\cite{Wim2005}.
This view is corroborated by numerical simulations (Fig.\ \ref{simulation}, upper panel) and experimental observations \cite{Zenesini:2009}.
This time evolution is due to the fact that the coupling between
the first and the second band is maximal at the edge of the first
Brillouin zone, for $k=\pm\pi/d_L$, and thus significant
transitions occur there, with periodicity $T_{\rm B}$. Figure \ref{simulation} shows that
plateaus are clearly present for $V_0 =1.5$ (shallow lattice, upper panel), but start to wash out for $V_0 =4.5$ (lower panel).
The range of validity of the plateau picture is further discussed in Appendix\ 
\ref{shallowlatt} and is approximately valid for $V_{0} \lesssim 4.5$.
In the following analysis we shall focus on this regime.

The approximated dynamics takes into account experimental and
numerical evidence and is valid for small values of $V_0$ and
$F_0$, when the transition times can be considered much smaller
than $T_{\rm B}$. We assume that the evolution inside the first band is
adiabatic for all $k$, except for $k \simeq \pi/d_L$, when a
transition towards the state with the same quasi-momentum in the
second band becomes possible. 
This transition will be effectively described by the evolution operator of the form
\begin{equation}
\tilde{U}=\left(\begin{array}{cc} s_{12} & -p_{12} \\ p_{12} &
s_{12} 
\end{array}\right) \,,
\end{equation}
with $p_{12}=\sqrt{1-s_{12}^2}$.
The operator $\tilde{U}$ acts on the two-dimensional space spanned
by $\{\ket{1},\ket{2}\}$, where $\ket{1}$ represents the state with $k=\pi/d_L$ in
the first band and $\ket{2}$ the state with same quasi-momentum in
the second band.

The transition from the second to the third band can be
schematized as the loss of a fraction $1-s_{23}^2$ in the population
of the second band towards a continuum, occurring at the crossing
around $k=0$. This assumption is justified for small values of
$V_0$ (see discussion above), such that a particle in the third (or higher) band
can be considered free.

\begin{figure}
\begin{center}
\includegraphics[width=0.5\textwidth]{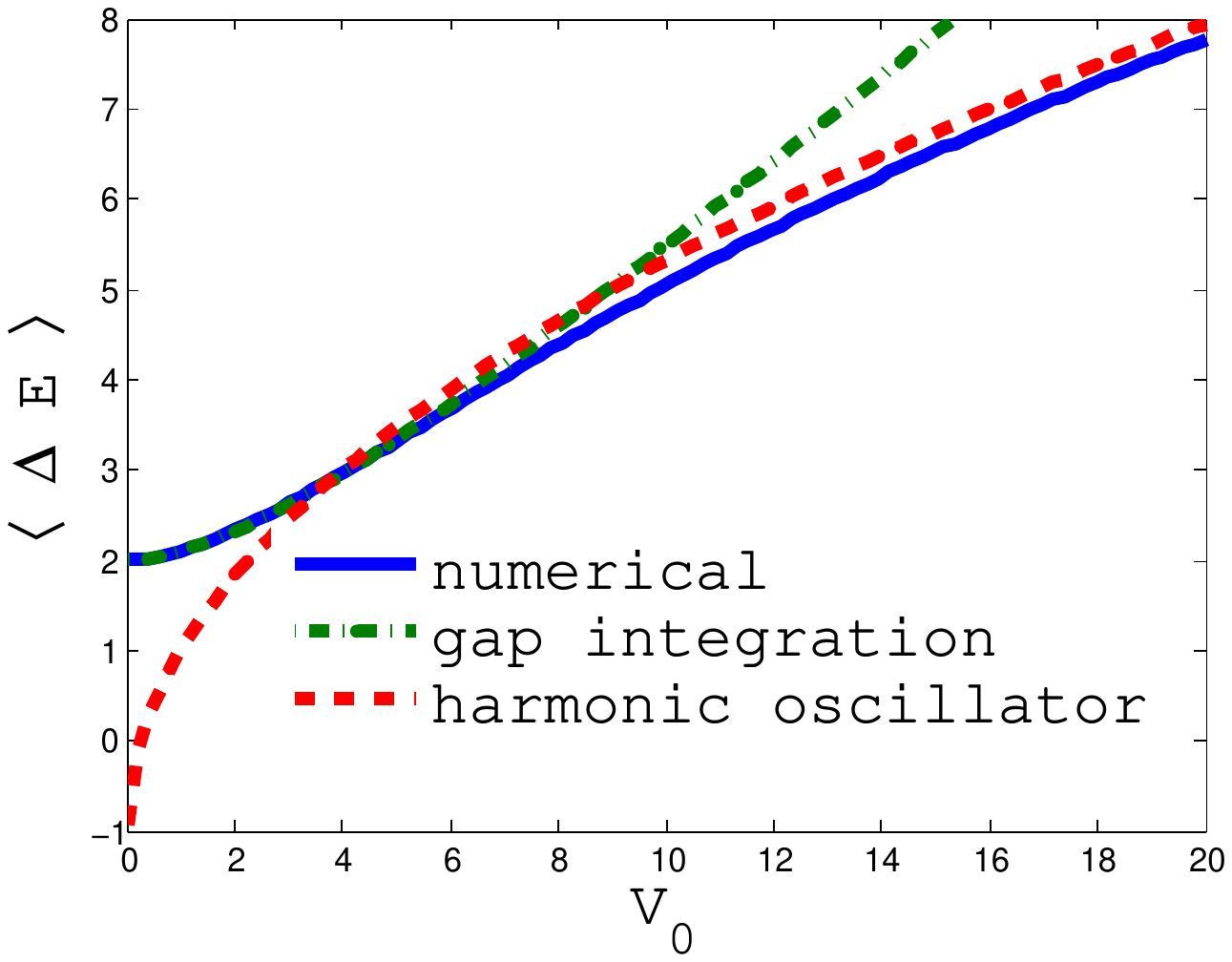}
\caption{ (color online)
Average band gap $\langle \Delta E \rangle$ vs $V_0$, both in $E_{\mathrm{rec}}$ units.
Comparison between numerical results, analytical results from Eq.\ (\ref{phiLZ}), 
and the harmonic oscillator approximation (\ref{phiLZlarge}).
In the small $V$ regime the band integration yields a good approximation, while for larger $V$, where the coupling becomes continuous, the harmonic oscillator approximation is more effective.
}
\label{cogap}
\end{center}
\end{figure}

During each Bloch cycle separating two successive transitions, the
relative phase between the second and the first band amplitudes
increases by (\ref{Blochphase}), which reads
\begin{equation}
\phi(V_0,F_0)=\frac{2\pi}{F_0} \langle \Delta E(V_0)
\rangle,
\label{phi12}
\end{equation}
where $\langle \Delta E \rangle$ is the energy difference
(in units $E_{\mathrm{rec}}$) between the second and the first
band, averaged over $\mathcal{B}$. This quantity can be exactly computed by
using Mathieu characteristics $a(\kappa,q)$, which are the
eigenvalues of the Mathieu equation \cite{Poole}
\begin{equation}
\frac{d^2y}{dx^2}+\left[ a - 2q \cos(2x) \right]y=0 ,
\end{equation}
corresponding to the Floquet solutions $y(x)=\exp(i\kappa x)
u(x)$. For small $V_0$, a good estimate is given by a Landau-Zener
gap integration
\begin{equation}
\label{phiLZ}
\langle \Delta E \rangle \simeq  \frac{1}{4}
\sqrt{64+V_0^2} + \frac{V_0^2}{32}\mathrm{arcsinh} \frac{8}{V_0}.
\end{equation}
For larger $V_0$, a tight-binding, or harmonic oscillator,  approximation yields
\begin{equation}
\label{phiLZlarge}
\langle \Delta E \rangle \simeq \sqrt{4V_0}-1.
\end{equation}
The exact result and the two aforementioned
approximations are compared in Fig.\ \ref{cogap}.

The effects of the dynamics in a time $T_{\rm B}$ from one transition to
the next one can thus be modelled in the basis
$\{\ket{1},\ket{2}\}$ by an effective non-unitary operator
\begin{equation}\label{opW}
W=\left(
\begin{array}{cc}
1 & 0 \\ 0 & s_{23}\mathrm{e}^{i\phi}
\end{array} \right) .
\end{equation}

By making use of this simplified model, we describe the time evolution in the following way.
At $t=0$ the condensate is in the first band, with quasi-momentum close to
$k=0$. As the lattice is accelerated, the quasi-momentum increases
until it reaches $\pi/d_L$ at $t=T_{\rm B}/2$, where the operator
$\tilde{U}$ comes into play and transfers part of the population to the
second band. The evolution from $T_{\rm B}/2$ to $3T_{\rm B}/2$ is summarized
by the application of $W$. Then, the second transition occurs, and
part of the population in the second band (decreased by losses towards the third band) can tunnel back
to the first band due to the action of $\tilde{U}$ and gives rise
to interference effects. The same steps occur in the subsequent
transitions.

On a time span $T_{\rm B}$, the dynamics of the system is therefore determined by
the successive action of the non-unitary operator
\begin{equation}\label{evored}
U=\tilde{U}W=\left( \begin{array}{cc} s_{12} &
-p_{12}s_{23}\mathrm{e}^{i\phi} \\ p_{12} & s_{12}
s_{23}\mathrm{e}^{i\phi}
\end{array}\right)
\end{equation}
in the basis $\{ \ket{1},\ket{2} \}$. The order of the two
operations is not relevant, since $W$ acts trivially on the
``initial state'' $\ket{1}$ before the first transition.

Besides the phase $\phi$, the operator $U$ depends on two other
independent parameters, namely the survival amplitudes $s_{12}$ and
$s_{23}$. $s_{12}$ represent the survival
amplitude in the first band after the {\it first} transition,
which is in fact comparable to a LZ process since the second band
is initially empty. The survival probability $s_{23}$ is related
to a LZ tunneling from the second to the third band, if we assume
the third band to be empty before each transition process.
A graphical representation of the parameters appearing in
Eq.\ (\ref{evored}) is given in Fig. \ref{anschaulich}.

Using the LZ critical acceleration for the first and second band
gap \cite{Landau, Zener, Niu2, Fishman}, analytical expressions
for $s_{12}$ and $s_{23}$ as functions of the microscopic
parameters can be obtained. At lowest order in $V_0$, the
survival amplitudes read
\begin{eqnarray}
\label{LZamplitudes}
s_{12}(V_0,F_0) &= & \sqrt{1-P_{\mathrm{LZ}}^{(1,2)}(V_0,F_0)}
\nonumber
\\ &= & \sqrt{1-\exp\left(-\frac{\pi^2 V_0^2}{32 F_0}\right)},
\\ s_{23}(V_0,F_0) & = & \sqrt{1-P_{\mathrm{LZ}}^{(2,3)}(V_0,F_0)}
\nonumber
\\ & = & \sqrt{1-\exp\left(-\frac{\pi^2
V_0^4}{32\cdot 16^2 (2 F_0)}\right)},
\label{LZamplitudes2}
\end{eqnarray}
where $P_{\mathrm{LZ}}^{(i,j)}$ is the Landau-Zener transition probability (\ref{plz}) from band $i$ to band $j$.

The evolution on a timescale $T_{\rm B}$, determined by a sequence of
$U$ operations, will be analyzed in detail in the following section.

\section{Transient and asymptotic behavior}
\label{intuitive}
We now specialize the model outlined in Section \ref{study} to the Pisa experimental setup \cite{Zenesini:2009,Tayebirad:2010}.
The state of the system
before the first transition is $\ket{1}$. Immediately after the
$n$-th transition, occurring at time $t=T_{\rm B}(n+1/2)$, the state of
the system is
\begin{equation}\label{evolved}
\ket{\Phi_n}=U^n\ket{1}.
\end{equation}
The matrix $U$ in Eq.\ (\ref{evored}) can be diagonalized, yielding
eigenvalues $(e_1,e_2)$. By expanding the initial state as
\begin{equation}\label{initial}
\ket{1}=c_1\ket{\psi_1}+c_2\ket{\psi_2},
\end{equation}
where $\ket{\psi_{1,2}}$ are the normalized {\it non-orthogonal}
eigenvectors of $U$, the state of the system at time $T_{\rm B}(n+1/2)$
is
\begin{equation}\label{evolution}
\ket{\Phi_n}=c_1 e_1^n \ket{\psi_1} + c_2 e_2^n \ket{\psi_2}.
\end{equation}
Due to the dissipative term in $W$, the two eigenvalues are
smaller than unity, and one of them, say $e_1$, is larger in
modulus than the other one. Thus, for $n$ sufficiently large, the
evolution reaches an asymptotic regime, in which the state after
the $n$-th transition is determined only by the state after the
previous one, with a transition rate depending on the largest
eigenvalue. Since the survival probability in the first band can
be defined as $P_n=|\langle 1 | \Phi_n \rangle|^2$, in the
asymptotic regime one gets
\begin{equation}
P_n\simeq |e_1|^2 P_{n-1}.
\end{equation}
By defining an asymptotic transition rate
\begin{equation}
\gamma =-\log \left( |e_1|^2 \right),
\end{equation}
it is possible to introduce a function $P_{\rm Z}(t)$ that
coincides with the value of the survival probability at the center of the
plateaus, at times $t=nT_{\rm B}$:
\begin{equation}
P_{\rm Z}(t)= Z \exp \left( -\gamma t \right).
\label{expfit}
\end{equation}
Compare with Eq.\ \eqref{Pexp}.
The parameter $Z$ in Eq.\ (\ref{expfit}) is in general
different from unity, due to the transient regime at the beginning
of the evolution. It represents the extrapolation of the
asymptotic exponential probability back at $t=0$. \\
\indent We now derive an analytical expression for $Z$. In the asymptotic regime, the system evolution described by  Eq.~(\ref{evolution}) corresponds to an evolution
operator applied to an initial unnormalized vector $\ket{\Psi_0}\equiv
c_1\ket{\psi_1}$:
\begin{equation}
\ket{\Phi_n} \simeq c_1 e_1^n\ket{\psi_1}=U^n\left(
c_1\ket{\psi_1}\right) = U^n \ket{\Psi_0}.
\end{equation}
The $Z$ parameter, representing the extrapolation of the
asymptotic behavior back to $t=0$, can be defined as the square
modulus of the projection of the fictitious initial vector $\ket{\Psi_0}$, onto the actual initial state $\ket{1}$
\begin{equation}\label{Zdef}
Z\equiv |\bra{1}\Psi_0\rangle|^2=|c_1|^2
|\bra{1}\psi_1\rangle|^2,
\end{equation}
which corresponds to an extrapolated ``survival probability'' in
the subspace spanned by $\ket{1}$, evaluated at the initial time.
$Z$ can be analytically computed as a function of the
independent parameters of the model, by explicitly diagonalizing
$U$. One obtains
\begin{widetext}
\begin{equation}\label{zeta}
Z(s_{12},s_{23},\phi)=\frac{ \left[
\frac{s_{12}}{2}(1-s_{23}\cos\phi)+\sqrt{\frac{\mathcal{K}(s_{12},s_{23},\phi)}{8}}
\right]^2 +s_{23}^2\sin^2\phi \left[
\frac{2-s_{12}^2(1+s_{23}\cos\phi)}{\sqrt{2\mathcal{K}(s_{12},s_{23},\phi)}}+\frac{s_{12}}{2}
\right]^2}{ \frac{\mathcal{K}(s_{12},s_{23},\phi)}{2}+
\frac{2s_{23}^2\sin^2\phi}{\mathcal{K}(s_{12},s_{23},\phi)} \left[
2-s_{12}^2(1+s_{23}\cos\phi) \right]^2 },
\end{equation}
with
\begin{eqnarray}
\mathcal{K}(s_{12},s_{23},\phi)& = & s_{12}^2\left(
1+2s_{23}\cos\phi+s_{23}^2\cos(2\phi) \right)-4s_{23}\cos\phi
\nonumber \\ & & + \sqrt{ s_{12}^4\left(
1+2s_{23}\cos\phi+s_{23}^2 \right)-8s_{23} s_{12}^2
\left(\cos\phi+2s_{23}+s_{23}^2\cos\phi\right)+16s_{23}^2 }.
\end{eqnarray}
\end{widetext}
In order to gain a qualitative understanding of the dependence of $Z$
(and $\gamma$) on the phase difference $\phi$ acquired
during a Bloch cycle, let us compare the first and second transitions. Let $P_0=1$ be the initial value of the survival probability
in the first band. After the first transition, the survival
probability becomes
\begin{equation}
P_1=s_{12}^2 P_0\equiv \mathrm{e}^{-\gamma_0}P_0.
\end{equation}
At the second transition, the discrepancy with the LZ prediction
becomes manifest. Since, in the parameter regime of small $V_0$ we
are considering, the ratio $s_{23}/s_{12}$ is very small [see Eqs.\
(\ref{LZamplitudes})-(\ref{LZamplitudes2})], we can apply a first-order approximation,
yielding
\begin{equation}
P_2\simeq (s_{12}^2 - 2 s_{23} p_{12}^2 \cos\phi) P_1\equiv
\mathrm{e}^{-\gamma_1}P_1.
\end{equation}
Thus, if the phase is  $\phi=2\pi j$, with $j\in\mathbb{Z}$, the second
transition is enhanced with respect to the first one. In this
case, a local maximum in the transition rate as a function of
$F_0$ is expected. On the contrary, if $\phi=(2j+1)\pi$, the
second transition is less pronounced than the first one.
\begin{figure}
\includegraphics[width=0.5\textwidth]{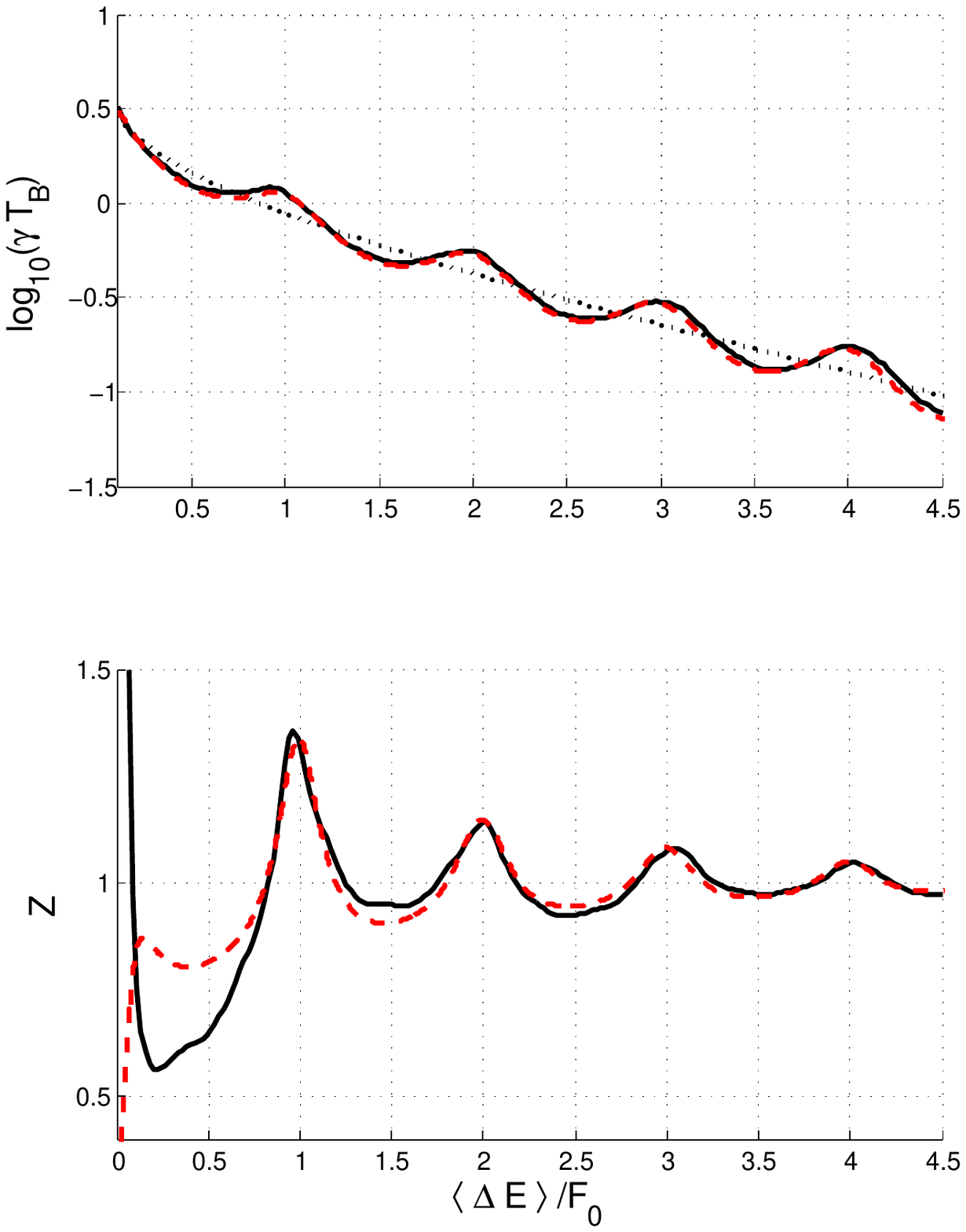}
\includegraphics[width=0.5\textwidth]{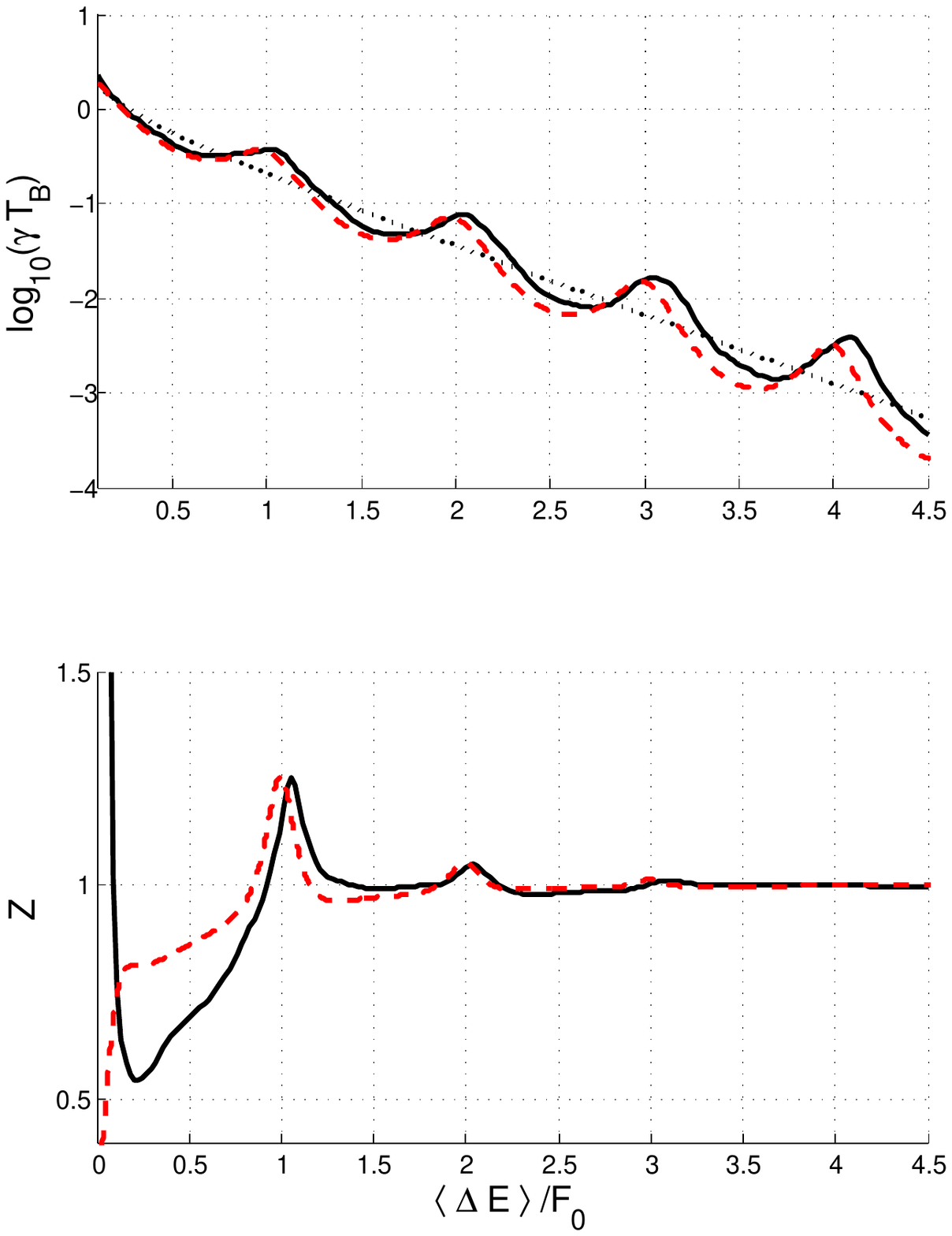}
 \caption{(color online)
Decay rate $\gamma$ and wave-function renormalization $Z$ vs $\langle \Delta E \rangle /F_0=\phi/2\pi$. Comparison among analytical results, obtained by exact diagonalization of the
reduced evolution operator $U$ in Eq.\ (\ref{evored}) [(red) dashed
lines], numerical simulations based on Eq.~(\ref{Hamiltonianequation}) (solid lines) and
Landau-Zener prediction  (dotted lines for $\gamma$). Upper two panels: $V_0=2$; lower two panels: $V_0=4$. }
\label{analyticalZZ}
\end{figure}

A backwards extrapolation of the second step gives a rough
estimate of the $Z$ parameter, which we call
$Z_1$:
\begin{equation}\label{Z1}
Z\simeq
Z_1=\mathrm{e}^{\gamma_1-\gamma_0}\simeq 1+ 2 s_{23}
\left(\frac{p_{12}}{s_{12}}\right)^2 \cos\phi.
\end{equation}
Even if Eq.\ (\ref{Z1}) represents a rather crude approximation, it
brings to light the correspondence between resonances in the
asymptotic transition rates and resonances in the $Z$
parameter. Quantities like (\ref{Z1}) are very useful in an
experimental context, where only the first few steps in the Bloch
cycles are accessible. If the survival amplitude can be measured
up to the $N$-th transition, the $Z$ parameter can be
approximated by
\begin{equation}
\label{ZN}
Z \simeq Z_N=
\mathrm{e}^{N\gamma_N-\sum_{n=0}^{N-1} \gamma_n}.
\end{equation}
At the same time,
\begin{equation}
\label{gN}
\gamma \simeq \gamma_N.
\end{equation}
The convergence to the real value of $Z$ is typically very
fast, and the first few cycles are already sufficient to obtain an excellent
approximation.

The estimates of Eqs.\ (\ref{phi12})-(\ref{LZamplitudes}), together with
Eq.\ (\ref{zeta}) enable one to obtain an analytical expression
$Z(V_0,F_0)$, yielding the value of $Z$ as a
function of the microscopic parameters.
Figure \ref{analyticalZZ} shows a comparison of the numerical
calculation and the estimates for $\gamma$ and
$Z$ with our analytical model. It is clear that
the model yields a better approximation for smaller $V_0$.
For $V_0 \gtrsim 4.5$ the peaks of $Z$ are
overestimated and the picture of successive tunneling events with
an intermediate phase accumulation becomes less valid. In the regime of
small $V_0$, the analytical model is very efficient, as long as $F_0$ is
not too large and the LZ tunneling rates do not have to be
adjusted due to the finite initial time of the evolution
\cite{Holthaus}.

\section{Experimental configurations}
\label{exptconf}

This section contains a discussion of the experiments performed up to now and suggestions for future measurements aimed at controlling the decay by a manipulation of the phase of the temporally evolved atomic wave packet.
The relations of Sec.\ \ref{intuitive} can be tested experimentally as follows.

\subsection {\bf Measurement of $P(t)$}

An experimental  check of the theory at the basis of the wave-function renormalization $Z$  is obtained by measuring the survival probability $P(t)$ for a time up to five Bloch periods for different parameter values, as in Fig.\ \ref{simulation}, and then introducing a fit with the exponential law of Eq.~\eqref{expfit} for the survival probability  at times $t=nT_{\rm B}$. The $Z$ and $\gamma$ parameters are determined by such a fit. The results of this approach are discussed in the following for the case of a narrow atomic momentum distribution, as in the RET experiments at Pisa with a Bose-Einstein condensate~\cite{Tayebirad:2010,Sias:2007,Zenesini:2008,Zenesini:2009}, and for the case of a broad atomic momentum distribution, as for the experiment performed at Austin \cite{Wilkinson:1997,Fischer:2001}.

\subsubsection{Pisa RET experiment}

\begin{figure}
\includegraphics[scale=0.5]{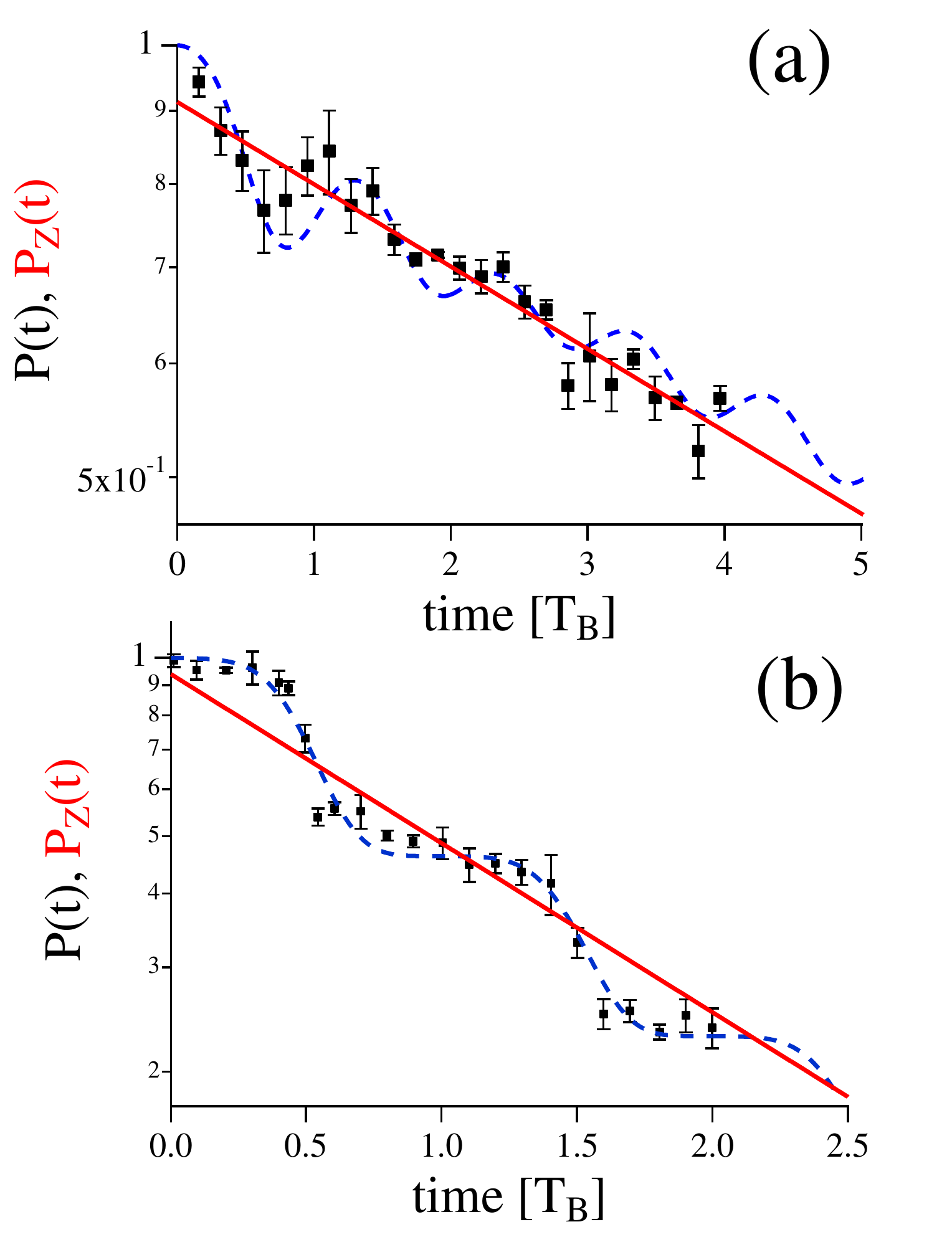}
\caption{(color online) $P(t)$: experimental results (squares) and  numerical solution of the Schr\"odinger equation describing the atomic cloud within the accelerated optical lattice (blue dashed line). The (red) continuous lines are exponential fits to the experimental data based on $P_{\rm Z}(t)$ by Eq.~\eqref{expfit}, whose crossing  with the $y$ axis yields the value of $Z$.  In (a) $V_0 = 5.8 E_{\textrm{rec}}$,
$F_0 = 5$ and in (b)  $V_0 = 1 E_{\textrm{rec}}$, $F_0= 0.383$. Both cases yield $Z<1$. The slope of the exponential decay gives the decay rate $\gamma$. }
\label{SurvivalProbability}
\end{figure}

\begin{figure}
\includegraphics[scale=0.5]{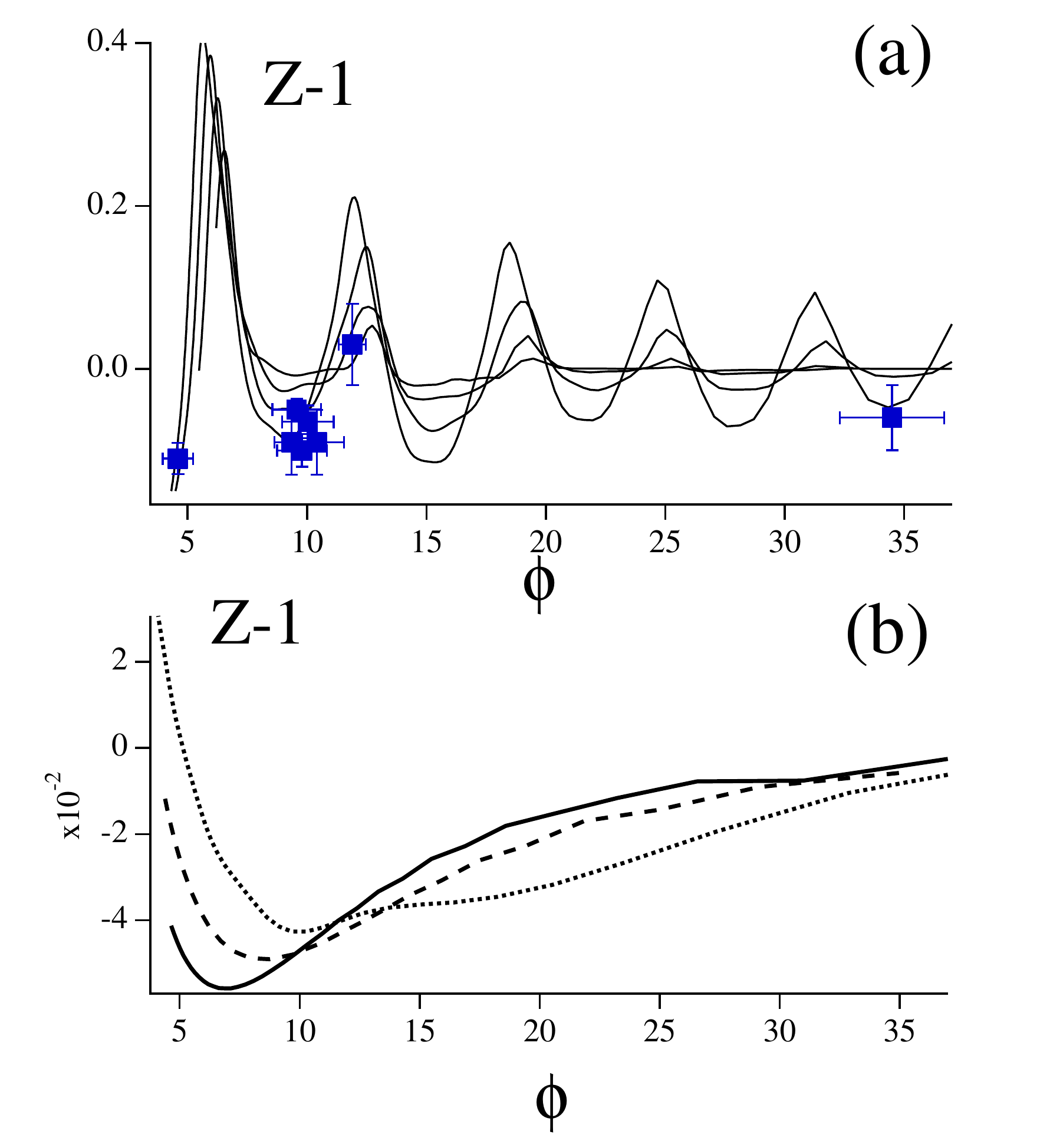}
\caption{(a) Scaling plot of $Z-1$ vs.\ $\phi$ of Eq.\ \eqref{phi12}, derived from RET experimental results (squares) using a narrow atomic quasi-momentum distribution. The experimental point at $\phi=4.8$ is obtained from the data of Fig.\ \ref{SurvivalProbability}(a) and the point at $\phi=34$ from the data of Fig.\ \ref{SurvivalProbability}(b). Full lines are the theoretical predictions for $V_0=1,2,3,4$. The RET coupling yields the oscillating behavior of $Z$ vs $\phi$, with the oscillation amplitudes increasing at lower $V_0$ for a fixed $\phi$. (b) Theoretical prediction for $Z-1$ in an Austin-type experiment, with a broad atomic quasi-momentum distribution, at $V_0=3,3.5$ and $4$ (continuous, dashed and dotted line, respectively). }
\label{RETScaling}
\end{figure}

The time dependence of the adiabatic survival probability was measured by freezing the tunneling process through projective quantum measurements on the states of the adiabatic Hamiltonian~\cite{Zenesini:2009}. Experimental results of $P(t)$ for different values of the lattice depth and the applied force are shown in Fig.~\ref{SurvivalProbability}. The solid and dashed lines are a numerical simulation of our experimental protocol and an exponential decay fit for our system's parameters, respectively. The vertical intercept of the exponential decay at $t=0$ gives the value of and the exponential decay rate gives the value of $\gamma$.\\
 The resonant tunneling appears as a strong variation for the exponential decay rate of $\gamma$ as a function of $\phi$, as measured in the experiments \cite{Sias:2007, Zenesini:2008}. This variation matches the numerical predictions of Fig.\ \ref{analyticalZZ}.\\
\indent Measured values of the $Z$ parameter vs the $\phi$ parameter are plotted in Fig.\ \ref{RETScaling}(a).  The error bars on the $Z$ values are determined by the exponential fits, as in Fig. \ref{SurvivalProbability}. Notice that $Z$ values both larger and smaller than one are measured. The error of the phase $\phi$ is linked to the experimental accuracy of the $V_0$ and $F_0$ parameters ($V_0$ carries an error of around ten percent). The experimental results are compared to theoretical predictions for the numerical solutions of the time-dependent adiabatic survival probability. The peaks in the plot are determined by  RET resonances. The simulation of Fig.~\ref{analyticalZZ} evidences that the dependence of $Z$ on $\phi$ matches the dependence  of $\gamma$. The position of the largest peak corresponds to the main  resonance \cite{Sias:2007,Zenesini:2008} $\Delta i=1$, and the positions of the smaller peaks are in agreement with those of higher order resonances.  The agreement between the theoretical and experimental determinations of $Z$ is very good, taking in account the difficulties of a precise determination of the lattice depth $V_0$. It should be noticed a posteriori that the experimental results are more easily produced in the case $Z<1$.

\subsubsection{Austin experiment on non-exponential decay}

The very broad atomic distribution of the experiment perfomed by Raizen's group in Texas \cite{Wilkinson:1997,Fischer:2001}, occupying several Brillouin zones, leads to a different temporal evolution of the survival probability. In particular, the deeper lattice potentials used in these works imply a different behavior of the $Z$ function. The survival probability was numerically evaluated on the basis of the theoretical treatment reported in Niu and Raizen~\cite{Niu:1998} and Wilkinson {\em et al.}~\cite{Wilkinson:1997}.  For the case of Rb atoms and parameters very close to those experimentally investigated in Pisa, Fig.\ \ref{RETScaling}(b) reports the $Z$ function versus the parameter $\phi$ at a fixed value of the
lattice depth. It may be noticed that the values of $|Z-1|$ are smaller than those measured in the case of a narrow atomic quasi-momentum distribution. The $Z$ dependence on $F_0$ is very smooth, without the oscillations of  Fig.\ \ref{RETScaling}(a).  The Niu-Raizen theory \cite{Niu:1998,Wilkinson:1997} includes only the two lowest energy bands and does not take into account tunneling phenomena such as RET or higher excited energy bands. The Niu-Raizen model is thus essentially a two-state model for Landau-Zener coupling, neglecting resonant tunneling effects, and averaged over all quasi-momenta in the entire Brillouin zone. Such a model is better suited for large values of $V_0$, when the energy bands become flat.

\subsection{Phase control}
\label{phasecontrol}
To further verify that the phase $\phi$ is, indeed, the important quantity determining the temporal evolution of the atomic wave function, it could be interesting to perform a LZ experiment for which the atomic acceleration is stopped after each Bloch period for a time $t_{\mathrm{halt}}=\pi / \Delta E$, with $\Delta E$ the energy difference between the two bands,  in order  to reverse the phase of the wave function's evolution. Differences in the predicted time dependence of $P(t)$  with and without this phase reversal are reported in Fig.~\ref{simulation}. Even if the experimental error introduced by the phase imprinting could be too large to derive $Z$ precisely in this regime, the observation of a modified decay rate in the presence of a phase reversal would represent a direct proof that $\phi$ is responsible for the resonances in the decay rate.

The survival probability obtained in an experiment where after each period  one halts or does not halt, with equal probability, represents another tool for modifying and testing the interference in successive Landau-Zener processes. The change of the decay rate by this randomization is equivalent to the change that would be obtained via \emph{bona fide} quantum measurements, as in the standard formulation of the Zeno effect which was experimentally oberved in \cite{Fischer:2001}. It can be demonstrated that the same atomic evolution is obtained by performing non-destructive survival probability measurements after each Bloch period, the quantum Zeno effect being achieved in the limit of very  frequent measurements carried out within a Bloch period.

\subsection{Emptying the second band}

A similar interesting experimental configuration is realized by totally eliminating the second band's occupation after each Bloch period. This could be produced as in the measurement protocol used in Ref.\ \cite{Zenesini:2009}, by decreasing the acceleration after each tunneling event from the ground band down to a small value such that the population in the second band tunnels to the continuum and is not confined anymore by the optical lattice. At the
same time  the population in the lower band does not tunnel to the second one, and is ready to be accelerated once again with the original large value. In this kind of setup all Landau-Zener steps in the survival probability as a function of time would have the same height on a logarithmic scale, determined by $s_{12}$ only. The phase $\phi$ would then be totally irrelevant for the atomic evolution.

\subsection{Links with quantum field theory}

Finally, from a theoretical perspective, it would be of great interest to explore the links with wave-function renormalization effects in quantum field theory. In that context, the quantity $Z$ arises from an analysis of the propagator
(enforcing probability conservation in the K\"all\'en-Lehmann representation \cite{KL,KL2})
and differs from unity at second order in the coupling constant. $Z$ is smaller than unity for stable states, but is unconstrained
and can become $>1$ for an unstable state.
There have been a few attempts
\cite{joichimatsu,hydrovanH,QFTC,BMT,AESG,giacosapagliara1,giacosapagliara2}
to analyze the quantum Zeno effect in the decay of elementary particles, but no experiment has been performed so far. It would be interesting to try and mimic these effects by making use of RET in BECs. This would take us into the  realm of quantum simulations.

\section{Conclusion and Outlook}
\label{concl}
In the pioneering work by Raizen \emph{et al.} \cite{Wilkinson:1997,Fischer:2001} the focus was on the deviations from exponential decay and the occurrence of the quantum Zeno effect and its inverse \cite{FP,KK}
due to repeated measurements.
In the present article we endeavored to go further and studied Landau-Zener transitions \cite{Landau,Zener}
under very different physical conditions, both in terms of initial state and parameters. 
This enabled us to use these effects as a
benchtest for the study of wave-function renormalization effects in quantum mechanics.
We have seen that by scrutinizing the features of the survival probability of the wave function that collectively
describes an ultra-cold atomic cloud, one can consistently define $Z$ and
extract crucial information on its behavior. It is remarkable that $Z$ can be directly measured and that its deviation from unity yields directly measurable consequences on the experimental observables. In addition, as the experimental
parameters are varied, $Z$ takes values that can be smaller or larger than unity.
If $Z<1$, the decay can be slowed down (quantum Zeno effect) or enhanced
(anti- or inverse-Zeno effect), but if $Z>1$, only the quantum Zeno effect is possible \cite{Pascazio}.

Our analysis of the atomic evolution in terms of successive free evolutions and
tunneling processes, with interference in the population occupations, points out that Landau-Zener
 transitions and St\"uckelberg oscillations
\cite{Stueckelberg} are two facets (one could say particular cases) of
the very complex problem of the atomic evolution within the periodic potential produced by the optical lattice, in
analogy to a previous analysis by Kling {\em et al.}~\cite{Kling:2010}.

For the shallow lattice regime, we have established a relationship
between $\gamma$, $Z$ and $\phi=2 \pi \langle \Delta E(V_0) \rangle / F_0$.
We have  demonstrated
that the Zeno regime and resonantly enhanced tunneling are both
controlled by the same parameter $\phi$ in an ultra-cold atomic cloud. The resonances in $Z$ can be explained by
a decay following the Landau-Zener probability in the first Bloch
period and resonantly enhanced decay in the following periods. In
contrast, the Niu-Raizen description \cite{Niu:1998} applied to describe the non-exponential decay of cold atoms in an optical lattice approximates the
tunneling rate from the second to the third band by one complete decay.
In the large $V_0$ parameter regime the RET resonances are not
important and do not affect the quantum Zeno effect. 


A future experiment could involve a BEC atomic cloud in the presence of atomic interactions \cite{niu,JonaLasinio:2003,Sias:2007,Zenesini:2008,AW2011,Wim2005,WimZ1,WimZ2,WimZ3,WimZ4}. As verified experimentally~\cite{JonaLasinio:2003}, in this case the tunneling probabilities are not symmetric ($s_{ij} \neq s_{ji}$) and the effect of the RET resonances could be enhanced or suppressed with attractive or repulsive interactions.

\begin{figure}
\begin{center}
\includegraphics[width=0.5\textwidth]{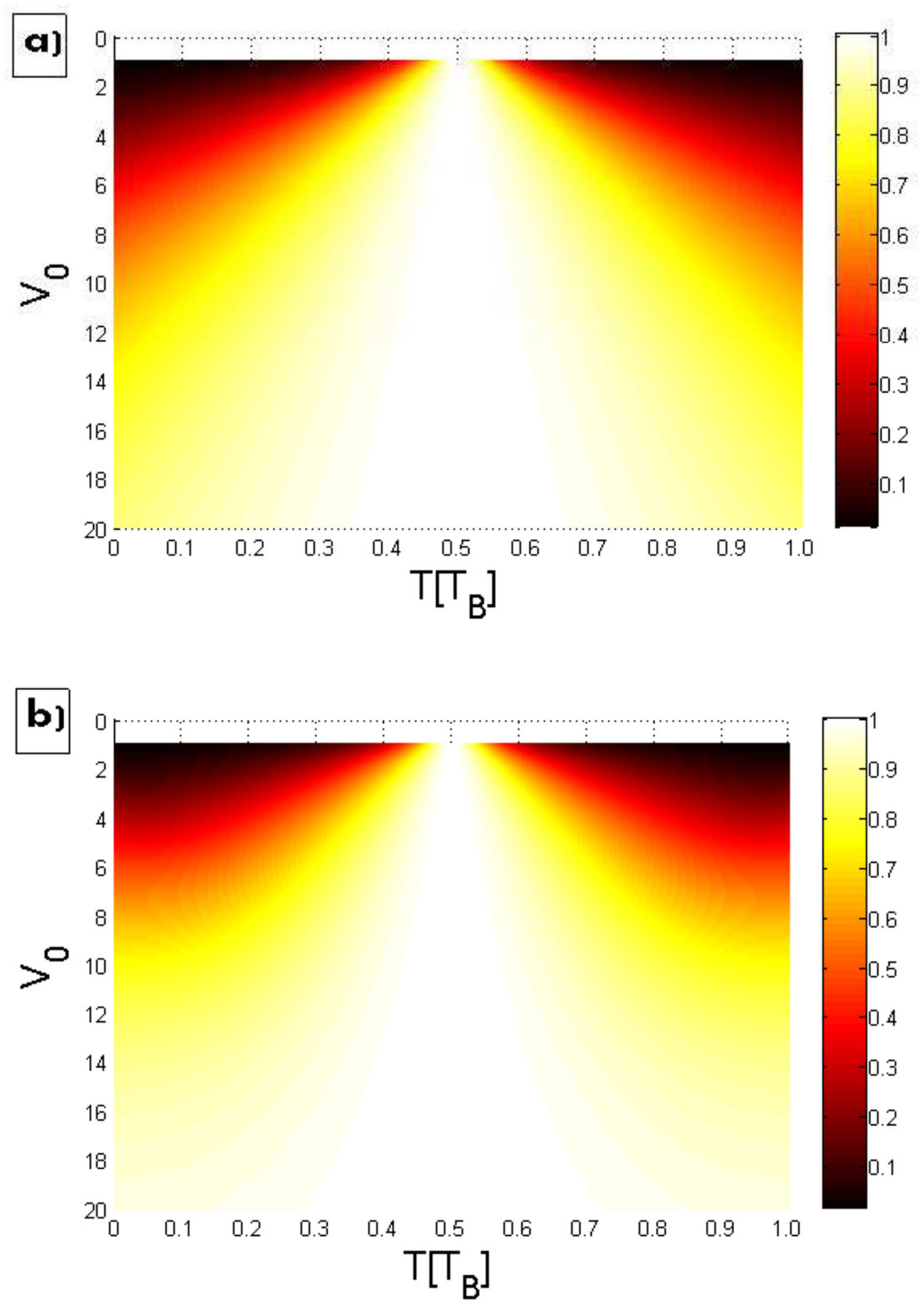}
\caption{(color online)  Adiabatic coupling strength $c(t)$ defined in Eq.~\eqref{coupling} and normalized to maximal coupling plotted vs. time and optical lattice depth.  Comparison between Lorentzian ansatz (upper panel) and numerical
results based on Eq.\ (\ref{Hamiltonianequation}) (lower panel).
The assumption of short tunneling events at the avoided crossings is valid for $V_{0} \lesssim 4.5$ (shallow lattice).
}
\label{FWHM}
\end{center}
\end{figure}

\appendix

\section {Check on the interrupted atomic evolution}
\label{shallowlatt}

The dynamics of interband tunneling is discussed in Sec.\ \ref{study} and hinges on the
assumption of a free phase evolution over the Brillouin zone, interrupted by a very short tunneling event at the avoided crossing, at well defined times $t=T_{\rm B}(n+1/2)$ with $n \in \mathbb{N}$, as in upper panel of  Fig.\ \ref{simulation} and in Fig.\ \ref{SurvivalProbability}(b).  To check the validity of this assumption we use the Hamiltonian $H_a$ which describes the time evolution in the adiabatic (energy) basis. $H_a$ can be obtained by expanding the  state $\ket{\psi(t)}$ of the system  in the time-dependent energy basis
\begin{equation}
\ket{\psi(t)}=\sum_n a_n(t) \ket{n(t)}
\end{equation}
and applying the Schr\"odinger equation $i \partial_t\ket{ \psi}=H \ket \psi$ with the Hamiltonian of Eq. \eqref{eqno8} to obtain
\begin{equation}
i\sum_n \left( \dot a_n \ket{n} +a_n \partial_t \ket{n} \right)=\sum_n a_n E_n \ket{n}.
\end{equation}
Taking the inner product with $\bra m$ and using $\braket {m|n}=\delta_{mn}$ we get
\begin{equation}
 \dot a_m  = -iE_m a_m -\sum_n \bra m \partial_t \ket{n} a_n
\end{equation}
and see that the off-diagonal term coupling the lowest two energy states is given by
\begin{equation}
\label{coupling}
c(t):=\bra 1 \partial_t \ket{2}.
\end{equation}
In the ideal Landau-Zener model of equation (\ref{landauzener}) and Ref.~\cite{Vitanov} this yields for $c(t)$ a Lorentzian function of time in a narrow time interval centered around the $T_{\rm B}/2$ transition time.
The  Lorentzian is displayed in Fig.\ \ref{FWHM}(a) for different values of the potential depth $V_0$. Figure
\ref{FWHM}(b) shows the numerical result for $c(t)$ in our system.
The model discussed in Sec.~\ref{study} ceases to be valid when $c$ is large, at the border of the Brillouin zone.
A comparison of the two plots in Fig.\ \ref{FWHM} clarifies that the approximations used in our analysis
break down for $V_{0}  \gtrsim 4.5$.

\bibliographystyle{apsrmp}

\end{document}